\documentclass[fleqn,usenatbib,usepdflscape]{mnras}
\pdfoutput=1
\usepackage{newtxtext,newtxmath}

\usepackage[T1]{fontenc}

\DeclareRobustCommand{\VAN}[3]{#2}
\let\VANthebibliography\thebibliography
\def\thebibliography{\DeclareRobustCommand{\VAN}[3]{##3}\VANthebibliography}

\usepackage{graphicx}	
\usepackage{amsmath}	
\usepackage[para]{threeparttable}

\def\beq{\begin{equation}}
\def\eeq{\end{equation}}
\def\(({\left (}
\def\)){\right )}
\def\H2{$\rm H_2$}
\def\cm1{$\rm cm^{-1}$}
\def\I{{\sc i}}
\def\II{{\sc ii}}
\def\Tc{$T_{\rm c}$}
\def\Te{$T_{\rm eff}$}

\title[Galactic chemical evolution]{Galactic chemical evolution of the 
solar neighborhood, solar twins and exoplanet indicators}

\author[C. R. Cowley, et al.]{
Charles R. Cowley,$^{1}$\thanks{E-mail: cowley@umich.edu (CRC)}
and Kutluay Y{\"u}ce$^{2}$
\\
$^{1}$Department of Astronomy, University of Michigan, Ann Arbor, MI 48109-1107, USA \\
$^{2}$Department of Astronomy and Space Sciences, Faculty of Science, 
University of Ankara, TR-06100 Tando\u{g}an, Ankara, Turkey \\
}

\date{Accepted 2022 mar 4; Received 2022.Feb 17; in original form 2022 Jan 11}

\pubyear{2015}

\hypersetup{draft}
\begin{document}
\label{firstpage}
\pagerange{\pageref{firstpage}--\pageref{lastpage}}
\maketitle

\begin{abstract}
Galactic chemical evolution (GCE), solar analogues or twins, and peculiarities of
the solar composition with respect to the twins are inextricably related.
We examine GCE parameters from the literature and present newly derived values
using a quadratic fit that gives zero for a Solar age (i.e., 4.6 Gyr). 
We show how the GCE parameters may be used not only to ``correct'' abundances
to the solar age, but to predict relative elemental 
abundances as a function of age. 
We address the question of whether the solar abundances
are depleted in refractories and enhanced in volatiles and
find that the answer is sensitive to the selection of a
representative standard.  The best quality data sets do not support 
the notion that the Sun is depleted in refractories and enhanced in
volatiles.   A simple model allows us to estimate
the amount of refractory-rich material missing from the Sun
or alternately added to the average solar twin.  The model
gives between zero and 1.4 earth masses.    

\end{abstract}



\section{Introduction\label{sec:intro}}

Galactic chemical evolution (GCE) 
refers to the relation between the abundances
of chemical elements and ages of the stars.  
Elemental abundances of the Sun and its twins are the 
keystone to galactic chemical
evolution in the solar neighborhood--a volume defined by stars with
mean distances of $\le$ some 100 pc \citep{bd18}.
The elemental abundance is typically
taken as the logarithmic ratio of an element to a reference element,
which is usually hydrogen or iron: $[El/H]$ or $[El/Fe]$. The bracket notation
indicates that the stellar ratio is relative to a standard.  That standard
is usually the Sun, although we explore the possibility of using some
average of nearby solar-type stars.  

In the small volume within some 100 pc of the Sun, secular changes due 
to stellar nucleosynthesis
are modified by the diffusion of stars from the interior and exterior of
the Galaxy \citep{man22}.  Possible stream remnants   
further complicate the picture \citep{mat21}, adding random or stochastic
components to any overall smooth relation
between mean abundance and age.  Nevertheless a number of studies have 
described the behavior of $[El/Fe]$ with stellar age by a simple linear
relation  (cf. \citet{bd18} and references below). 
We explore below the use of GCE parameters
which give zero for a star of the solar age. 

\section{The data}
\label{sec:data}
The present work is entirely based on published results of five studies
of stellar abundances and ages.  Three of these, \citet[][henceforth, BD18]{bd18}
which includes data from \citet{spin18},
\citet[][henceforth, NIS]{niss20}, and \citet[][henceforth, LIU]{liu20},
give data for fewer than 100 stars.  
Two, \citet[][henceforth, BR]{brew16} and \citet[][henceforth, DM]{dm17,dm19}, have data 
for more than 1000 stars.  Stellar ages have been taken directly from the 
cited references.   

All five works use the method of precision differential
abundances (PDAs, cf. \citet{mon13}.  
Abundance differences are based on ratios of abundances from
line pairs for each spectrum (e.g., Fe \I\ or Fe \II).
Individual stellar properties do enter, as model atmospheres
are used.  However, the abundance differences are more accurate
than individual abundances determined separately for each star, 
because uncertain factors such as oscillator
strengths cancel.  

Many papers relevant to PDAs, GCE, solar twins, and \Tc\ correlations
are not examined  here.   
A sample list includes:\
\citet{adib12}, \citet{adib14}, \citet{adib15},\citet{ben14},
\citet{bb16}, \citet{cas20}, \citet{gonz10}, \citet{gonz13},
\citet{ram09}, \citet{ram14a}, \citet{spin16b}, and \citet{ram14b}.  
The latter reference is of interest because it
lists all but one of the BD18 stars in a solar twin search.  
Additional references may be found in the 
exhaustive study of the solar twin problem by \citet{yana21}.

We focus here on the BD18 work for two reasons.  First, that study includes
30 elements from carbon to dysprosium, significantly more  elements than in the 
other studies considered here.  Second, we find the accuracy of
the abundances is at least as good as the best of the others. 
The NIS survey has 13 elements, C-Ni, Sr and Y. \,\,LIU has 17
elements C-Zn, Sr, Y, Zr, Ba-Dy.  
BR and DM are discussed below.

Tab.~\ref{tab:Abdifs} will
give some insight into the relative accuracy of the five studies.  We give 
averages for the absolute value of the pairwise abundance differences 
for carbon, iron, aluminium, and silicon for  stars in common.
The mean and/or standard deviations of the differences
in Tab.\ref{tab:Abdifs} indicate that uncertainties range from the
0.01 dex level to somewhat more than 0.1 dex for some elements.  
The BD18 and NIS differences are the smallest.


\begin{table}
	\centering
	\caption{Abundance absolute magnitude differences for 
    $[C/H], [Fe/H], [Al/H]$, and $[Si/H]$ from five studies\label{tab:Abdifs}}
	\begin{tabular}{l c c c c} 
    \hline
\multicolumn{5}{c}{|BD18-BR|} \\ \hline
         &  C     &  Fe    &  Al    &   Si     \\  \hline
Mean     & 0.044  &0.013   &  0.067 &  0.018   \\
St. Dev. & 0.032  &0.009   &  0.138 &  0.023   \\
No. Stars& 15     & 15     &   15   &   15    \\  \hline
\multicolumn{5}{c}{|BD18-DM|} \\ \hline
Mean     &        &0.112   &0.127   &  0.107  \\
St. Dev. &        &0.053   &0.093   &  0.020   \\
No. Stars&        &8       &8       &8    \\ \hline
\multicolumn{5}{c}{|LIU-NIS|} \\ \hline 
Mean     &0.073   &0.040   &0.080   & 0.025  \\
St. Dev. &0.084   &0.056   &0.085   & 0.025  \\
No. Stars& 3      & 4      &4       &4       \\ \hline
\multicolumn{5}{c}{|BD18-NIS|} \\ \hline
Mean     & 0.022  &0.007   &0.010   &0.008  \\
St. Dev. & 0.011  &0.005   &0.009   &0.005  \\
No. Stars& 22     &22      &22      &22    \\ \hline                   
\hline						
\end{tabular}
\begin{tablenotes}
\item No carbon abundances in DM
\item Carbon is an outlier for LIU-NIS in HD 4915 and is omitted
\end{tablenotes}
\end{table}
  
The BR and DM stellar samples are an order of magnitude more numerous 
(BR: 1615; DM: 1059) than BD18 and
significantly more diverse in \Te, $\log(g)$, and composition. 
The BR stars have 15 elements, including
nitrogen, missing from the other surveys.
DM also has 15 elements:  the
refractories Mg - Ti and the
intermediate volatiles Cu and Zn as well as Ba, Ce, Nd, and Eu. 

The relation between stellar abundance and age 
for these surveys is complex.  Plots
of abundance versus age reveal ``families'' of stars defined by
common spatial (thin or thick disk) or chemical (metal- or alpha-rich)
properties (see Fig. 7 of DM).  The BD18 data 
constitute a spatially and chemically more homogeneous group
than the larger samples.  Similar remarks apply to the NIS
and LIU data.

\section{Galactic Chemical Evolution--plots and parameters}
\label{sec:pandp}

Four of the five studies considered here, excepting  BR,
have obtained linear relations between $[El/Fe]$ and 
stellar age.  We report values of GCE parameters for the BR stars based
on a selected set of 160 stars described in Tab.~\ref{tab:slopes} (see
note to Col. 11). 

\begin{figure*}
	\includegraphics*[width=6 in]{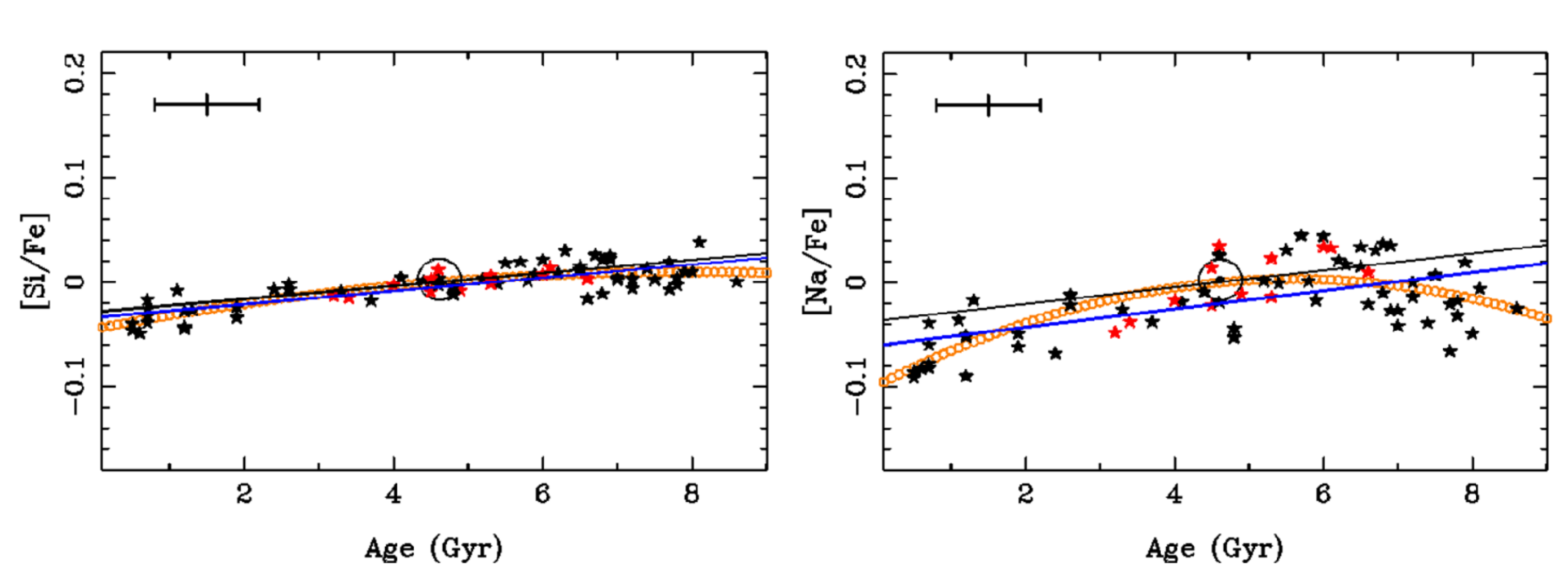}
    \caption{$[Si/Fe]$ and $[Na/Fe]$ vs. Age for 68 BD18 stars in Gyr.  
    Blue: unconstrained linear (UL) fit;
    Black: constrained linear (CL) fit; Orange: constrained quadratic (CQ) fit. In these
    plots, the standard abundance is not that of the Sun but 
    the average of the 12 BD18 twins of 
    table~\ref{tab:Dataset}.  Selected ``twins'' are plotted as red stars.  
    See text for further discussion.}
    \label{fig:one}
\end{figure*}

The most common GCE parameters are the slope and intercept of a linear
fit, or sometimes, a broken linear fit to a plots of $[El/Fe]$ vs. age
such as in Fig.~\ref{fig:one}.  

The elemental abundances of a star of a given age may considered as due
to smooth, secular changes (GCE) and other, peculiar, causes such as accretion 
from a nearby supernova or the ingestion of an earthlike planet.
These parameters may be used to compare the peculiar abundances of
stars that are not due to the secular effect of GCE.  If we want to compare
the peculiar abundances of solar twins to the Sun, we must remove
the differences due to GCE and obtain the abundances
the star would have (had) if its age were 4.6 Gyr, the age of the Sun \citep{bon02}.
We refer to these adjustments as ``corrections''.
Since the current linear fits do not give precisely zero corrections for 4.6 Gyr,
we have explored  
two relations that do give precisely zero for the solar age.  The first is a 
constrained linear (CL) relation
with only one parameter.  Its fits have  significantly larger variances than the 
two-parameter unconstrained linear (UL) fits; it is therefore not useful.  
The second is a constrained quadratic (CQ), which like the UL
fit has two free parameters.  

The CQ fit, as well as the broken linear fits, allow for a change
of slope with age.  This can be of importance when attempting to make a correction
for GCE, which depends on the slope of $[El/Fe]$ vs. age.  The corrections
add (or subtract) an amount from the $[El/Fe]$.  For example, the BD18 GCE
parameters, $m(age)$ and $b$ are used in
\begin{equation}
[El/Fe]_{\rm corrected} = [El/Fe]-[m(age)+b]
\label{eq:gcecor}
\end{equation} 
\noindent This linear correction
does not allow for a change in slope $m$, so if the true slope does change    
the "correction" could be in the wrong direction for some age range!
 
The GCE plots of silicon 
and sodium are shown in Fig.~\ref{fig:one}.  They represent the typical monotonic
behavior of the abundances with age, and the occasional element for which the
plot changes slope.  A liberal uncertainty estimate for the points in the
vertical direction is 0.022 dex, compatible with entry in Tab.~\ref{tab:Abdifs},
the mean of $|BD18-NIS|$ for carbon.  The symbol in the upper left of the diagrams
of Fig.~\ref{fig:one} represents these abundance uncertainties.    
The age uncertainty is taken as 1.4 Gyr
which is based on the standard deviation of independent age determinations of  NIS and \citet{spin18}.
We present similar plots for all 30 elements and 79 stars
of the BD18 study in the files PlotsofEloFe* 
at https://doi.org/10.5281/zenodo.6077735.  
 The archive contains both PowerPoint (pptx) and
pdf versions.

The CQ parameters are derived by least squares fits to plots such as those of Fig.~\ref{fig:one} 
and have the form
\begin{equation}
[El/Fe] = aa\cdot ({\rm age} - 4.6) + bb\cdot ({\rm age} - 4.6)^2,
\label{eq:abage}
\end{equation}
\noindent where $aa$ and $bb$ are constants for individual elements.  Age is in units of Gyr.  
The constant $aa$ is the slope at age 4.6 Gyr.

Values of the variance ratios for the UL and CQ fits are remarkably close
to unity for the BD18, NIS, and LIU samples.  Most have values
between 0.95 and 1.1, and none indicate a strong preference for the
UL or CQ fits.

Plots similar to those of Fig.~\ref{fig:one} were shown by \citet{dasil12} 
and \citet{spin16a}.  In the latter paper, three fitting techniques were
employed: a simple linear and a broken linear fit as well as a hyperbolic
fit (see their Fig. 1).  Most of those hyperbolic fits are in qualitative
agreement with those in PlotsofEloFe on zenodo: Na, Ca, Co, Ni, Cu, and
Zn. 
\begin{table}
	\centering
	\caption{Selected twins from BD18 stars and parameters}
	\label{tab:Dataset}
	\begin{tabular}{rrllrl} 
    \hline
HD      &HIP 	&T$_{\rm e}$ (K)&  $\log{g}$ & 	$[Fe/H]$&  age (Gyr) \\  
\hline
13357  &10175   &5719 &   4.485 &   -0.028  &3.2 \\
16008  &11915   &5769 &   4.48  &   -0.067  &3.4 \\
36152  &25670   &5760 &   4.42  &    0.054  &4.9 \\
88072  &49756   &5789 &   4.435 &    0.023  &4.5 \\
115169 &64713   &5788 &   4.435 &   -0.043  &5.3 \\
124523 &69645   &5751 &   4.435 &   -0.026  &5.3 \\
138573 &76114   &5740 &   4.41  &   -0.024  &6.6 \\
146233 &79672   &5808 &   4.44  &    0.041  &4.0   \\
145927 &79715   &5816 &   4.38  &   -0.037  &6.1 \\
167060 &89650   &5851 &   4.415 &   -0.015  &4.6 \\
183658 &95962   &5805 &   4.38  &    0.029  &6.0 \\
200633 &104045  &5826 &   4.41  &    0.051  &4.5 \\
{\bf BD18 stars} & {\bf averages}& {\bf 5785.}&  {\bf 4.427}  & {\bf -0.0035 }  &{\bf 4.86} \\
\hline						
\end{tabular}
\end{table}

\begin{figure*}
	\includegraphics*[width=6in]{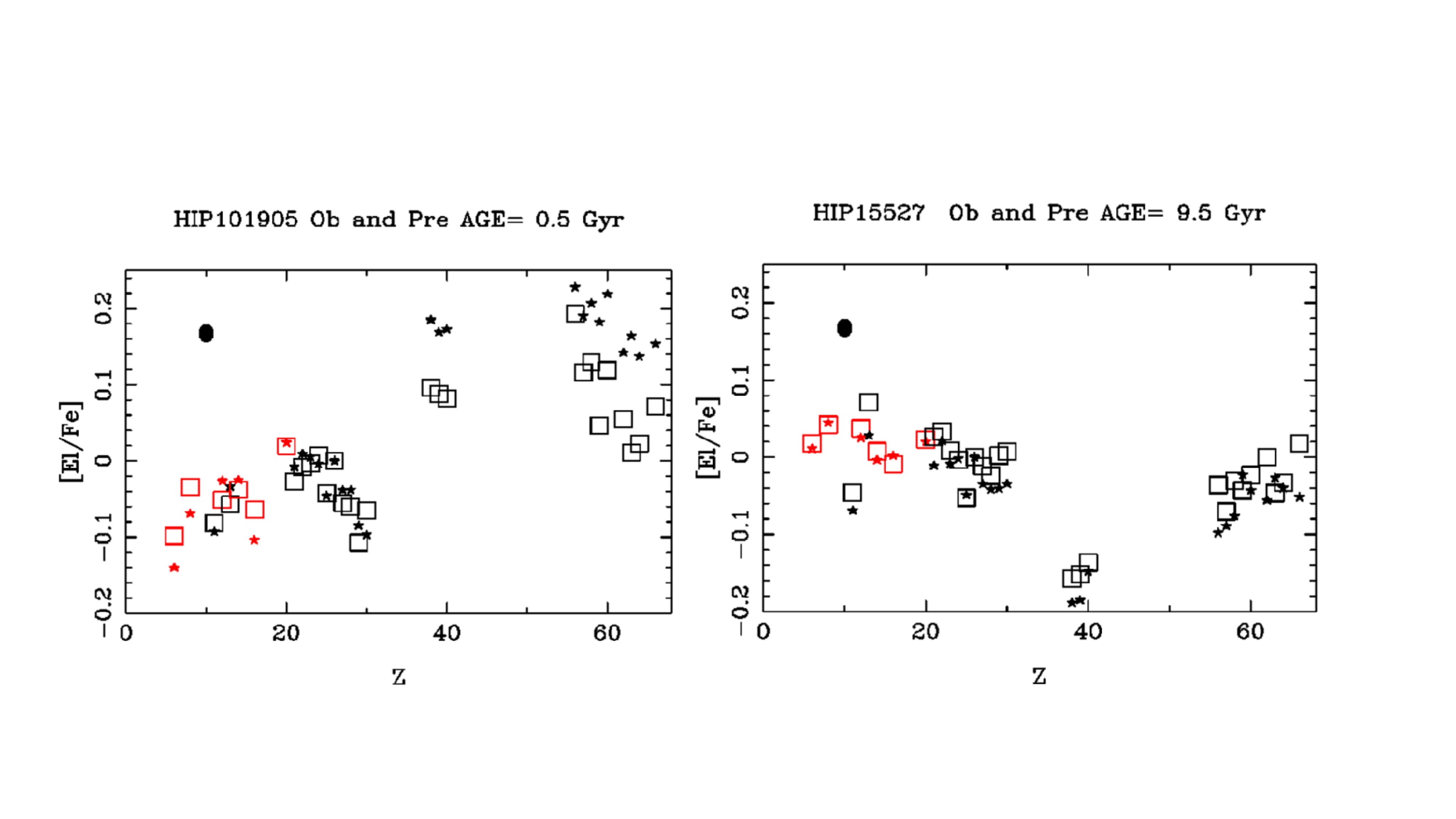}
    \caption{[El/Fe] vs. Z for 2 BD18 stars.  Stars represent observed abundances while
    squares come from the CQ ``average'' fits.  Alpha elements, C, O, Mg, Si, S,
    and Ca are shown in red.  The black mark in the upper-left corner represents the
    uncertainty in $[El/Fe]$ for the star points, roughly 0.02 dex.  The squares are
    computed directly from the CQ formula and do not have a similar uncertainty.
    See text for further discussion.}
    \label{fig:diffs}
\end{figure*}

NIS (see their Fig. 4) has plots of $[El/Fe]$ for 12 elements vs. age
similar to our Fig. 1.  The 
color coding of these plots emphasizes their conclusion that the sample
consists of two groups with distinct chemical history.  The first group
consists mostly of stars that are younger than the Sun, while stars of
the second group are mostly older.  This
is consistent with the morphology of our plots of $[El/H]$ vs. atomic number, $Z$,
which may be downloaded from zenodo.
An examination of those plots of the BD18 data shows that most of the stars
with positive overall shapes (POS) are younger than the Sun, while those with the V-shape
are older (see Fig.~\ref{fig:diffs}).

The overall shapes of the NIS plots (see their Fig. 4)
generally agree with our CQ fits, but not entirely.  For example,
the plot for $[Na/Fe]$ vs. age would support our negative curvature, except
for high outliers which are unusually sodium rich: HD 1461, HD 66428, 
HD 111031, HD 134487, HD 134606, HD 169691, and HD 204313.  These same
stars lie below the mean of the $[O/Fe]$-plot; alternately, they are
part of a separate population if broken GCE lines are adopted.

Figs. 4, 5, and 6 of LIU show $[El/Fe]$ vs age.  We examined similar plots
for their 68 ``comparison'' stars only, finding little convincing evidence
of curvature, even for Na and Cu. Only 4 of the LIU stars were in the 
BD18 and NIS surveys.

 
The larger BR and DM samples are  
broader in  the temperature-gravity spread. 
The DM stars are classified into populations.  Most, 882, are thin disk.
There are 108 thick disk, 60 high-$\alpha$, and 9 halo stars.  This mixture 
causes apparent clustering in certain age ranges, often somewhat displaced from
the general trend.  In addition to DM's published GCE parameters,   
we determined new UL ones based on severely limited
samples (see Tab.~\ref{tab:slopes} footnote to Col. 6).
The 73 DM stars had all been classified as thin disk.
BR does not explicitly classify its stars into populations, but notes the
occurrence of $\alpha$-rich stars at sub-solar metalicities.
Selection criterior for the BR stars are in the table footnotes
to Cols. 11 and 12.

We made GCE fits to $[El/Fe]$ vs. age for the BR and DM samples.  
Without careful sample selection, the
results show a considerable scatter due to the inclusion of populations that
were purged in the BD18 sample.  
Different populations are readily seen in DM's colorful plots
of $[El/Fe]$ vs age \citep[see][]{dm19}, Fig. 7.      

Unlike the other four surveys, BR concentrated on the techniques and 
results of abundance determinations;  GCE parameters were not obtained.  We
obtain them here for the limited sample of 160 BR stars.


In Tab.\ref{tab:slopes} we compare published and newly computed 
slopes of the GCE relations from
BD18, NIS, LIU and selections of stars from BR and DM.  All newly derived
parameters are from unweighted least squares fits.  All of the stars are
from the thin disk.
We followed BD18 and excluded 11 of their 79 stars that are old,
$\alpha$-rich or of the thick disk population.  Thus the plots 
of Fig.~\ref{fig:one} and in the zenodo archive are of the 68 thin disk
stars.  Similarly in \citet{niss15}'s study of 21 stars, they excluded 3 $\alpha$-enhanced
stars from their GCE fits.  In addition to the slopes from the 2015 paper, 
we present newly derived UL slopes based on the more recent \citet{niss20} results, 
one determined using 71 stars  excluding those with $[Fe/H] \ge 0.15$,
(Col. 8), and 40 (Col. 9) stars also excluding those with $|[Fe/H]|\ge 0.15$.
The \citet{liu20} figures are for the 68 ``comparison stars'' taken from that
paper.

New UL coefficients for 73 DM stars are based on the                                                                                                                                                                                                                                                                                                                                                                                                                                                                                                                                                                                                                                                                                                            
criteria specified in \ref{tab:slopes}, Col. 6.
The age accuracy condition was adopted from DM and also applies to Cols. 5 and 6 of
Tab.~\ref{tab:slopes}. 

UL and CQ coefficients were derived
in the present study from 160 stars of the BR study (Cols. 11 and 12). 
The 1615 BR stars were filtered as described in the supplement to Tab.~\ref{tab:slopes}.


The BR survey was the only one of the five discussed here to determine
an abundance for the element nitrogen.  This element is critical to the
search for correlations of $[El/Fe]$ with \Tc, as there are only
two other highly volatile elements that are spectroscopically available, 
carbon and oxygen.  The $[N/Fe]$
ratio was included in the extensive examination of carbon isotopes
and elemental nitrogen by \citet{botel20}.  Fig.~5 of that study shows $[C/Fe]$ and
$[N/Fe]$ vs. age; their broken-linear fit for nitrogen is in agreement
with our CQ fit to the BR data--positive slope for the younger stars,
and negative for the older.

Although the slopes of Tab.~\ref{tab:slopes} differ numerically, they generally agree as
summarized here:    
\begin{enumerate}
\item For the lighter elements, the slopes are positive or small (e.g. Ca). 
Positive slopes indicate an      
increase in the $[El/Fe]$-ratio for the older stars.
\item Beyond the iron-group (3d) elements the slopes are mostly negative,         
indicating an increase in $[El/Fe]$ for the youngest stars.
\item The slope parameter for exceptions to items (1) and (2), such as for Ca or Cr   
are small.
\item The triplet of neutron-addition (nA) elements, Sr, Y, and Zr has       %
very steep slopes in the BD18 data, and NIS data for Y.
\item The heaviest nA elements, from Ba to Dy all have negative      
slopes in the 68 BD18 stars.  The DM slopes from Ba to Dy are mixed.
  
\end{enumerate}

\begin{table*}
\begin{center}
\caption{GCE slopes for elements. CQ constants in Columns 2 and 3\label{tab:slopes}}
\end{center}
\begin{threeparttable}
\begin{tabular}{lrrrrrrrrrrr} \hline
 1  &    2        &    3     &     4   &    5     &    6     &   7     &   8     &   9     &   10     & 11 &12\\
el  &aa\rule{4mm}{0mm}& bb &BD18, m   & Fig.7(DM)&DM(UL) &   Nissen & Nissen  & Nissen  &  Liu    & Brew(CQ)&Brew(UL)      \\
\hline    
 C  &  0.01481 &-0.00227  & 0.0115 &  0.0187 &           &   0.0139 &   0.0162&  0.0147 &    0.013& 0.0091&  0.0084  \\
 N  &          &          &        &         &           &          &         &         &         & 0.0093&  0.0090  \\
 O  &  0.00849 & 0.00002  & 0.0088 &         &           &   0.0161 &   0.0201&  0.0172 &    0.002& 0.0141&  0.0137  \\
 Na &  0.00651 &-0.00326  & 0.0086 &         &           &   0.0098 &   0.0064&  0.0071 &    0.001& 0.0161&  0.0176          \\
 Mg &  0.01027 &-0.00054  & 0.0099 &  0.02059&   0.0149  &   0.0089 &   0.0126&    0.01 &    0.011& 0.0145&  0.0149  \\
 Al &  0.01424 & 0.00007  & 0.0139 &  0.0126 &   0.0113  &   0.0167 &   0.0168&  0.0163 &    0.033& 0.0255&  0.0272  \\
 Si &  0.00578 &-0.00086  & 0.0063 &  0.0075 &   0.0073  &   0.0006 &   0.0079&  0.0065 &    0.000& 0.0098&  0.0104  \\
 S  &  0.00757 &-0.00194  & 0.0098 &         &           &   0.0007 &         &         &    0.003&           \\
 Ca & -0.00045 & 0.00105  & -0.0011&  0.00616&  -0.0005  &  -0.0001 &   0.0026&  0.0002 &   -0.002& -0.0016& -0.0016  \\
 Sc &  0.00605 &-0.00015  & 0.0059 &         &           &          &         &         &    0.024&           \\
 Ti &  0.00418 & 0.00053  & 0.0036 &  0.01136&   0.0048  &   0.0063 &   0.0087&  0.0062 &    0.007&  0.0088& 0.0079  \\
 V  &   0.00122& 0.00010  &  0.0013&         &           &          &         &         &   -0.013&  0.0058& 0.0043  \\
 Cr &  -0.00131& 0.00012  & -0.0016&         &           &   -0.0026&  -0.0014&  -0.0018&    0.001& -0.0023& -0.0031  \\
 Mn &   0.00075&-0.00235  &  0.0023&         &           &          &         &         &   -0.012&  0.0066& 0.0072  \\
 Co &   0.00628&-0.00176  &  0.0074&         &           &          &         &         &    0.007&           \\
 Ni &   0.00575&-0.00218  &  0.0071&         &           &    0.0048&   0.0033&  0.0041 &    0.007&  0.0116& 0.0124  \\
 Cu &   0.01442&-0.00286  &  0.0149&  0.00609&   0.0112  &          &         &         &         &           \\
 Zn &   0.00932&-0.00161  &  0.0102&  0.01413&   0.0121  &   0.0117 &         &         &    0.027&           \\
 Sr &  -0.02749&-0.00096  & -0.0251& -0.01922&   -0.0255 &          &   -0.0172&-0.0211 &         &           \\
 Y  &  -0.02581&-0.00106  & -0.0238& -0.01658&   -0.0223 &   -0.033 &   -0.0207&-0.0233 &   -0.007& -0.0159&  -0.0149 \\
 Zr &  -0.02359&-0.00088  & -0.0219& -0.00787&   -0.0165 &          &          &        &         &           \\
 Ba &  -0.02898& 0.00441  & -0.0317& -0.01193&   -0.0196 &          &          &        &         &           \\
 La &  -0.02196& 0.00155  & -0.0227&         &           &          &          &        &         &           \\
 Ce &  -0.02007& 0.00282  & -0.0220& -0.00525&    -0.0079&          &   -0.0021&        &         &           \\
 Pr &  -0.01020& 0.00028  & -0.0103&         &           &          &          &        &         &           \\
 Nd &  -0.01800& 0.00271  & -0.0198&  0.00767&    -0.0021&          &          &        &         &           \\
 Sm &  -0.00727& 0.00149  & -0.0077&         &           &          &          &        &         &           \\
 Eu &  -0.00583&-0.00076  & -0.0056&  0.01058&    0.0085 &          &          &        &         &           \\
 Gd &  -0.00616&-0.00013  & -0.0060&         &           &          &          &        &         &           \\
 Dy &  -0.00780& 0.00234  & -0.0073&         &           &          &          &        &         &           \\
 \hline           
 \end{tabular}
 \begin{itemize} 
 \item (Col. 1) Element
 \item (Col. 2) $aa$, CQ slope at 4.6 Gyr, Eq. \ref{eq:aabb}, 68 stars, this work
 \item (Col. 3) $bb$, CQ constant, Eq. \ref{eq:aabb}, 68 stars, this work
 \item (Col. 4) UL slope $m(t)$ taken directly from BD18 Table 3, 68 stars 
 \item (Col. 5) DM GCE parameters a and b from Figure 7, for about 270 thin disk stars with age uncertainties $\le$ 1.5 Gyr 
 \item (Col. 6) 73 selected DM stars; age uncertainty $\le 1.5$ Gyr, age < 9 Gyr, 5480 K < \Te < 6080 K,
 4.24 < log($g$)< 4.64
 \item (Col. 7) Nissen et al. (2015) Table 6, 18 stars
 \item (Col. 8) Nissen et al. (2020), Vizier Cat., 71 stars
 \item (Col. 9) Same as 8, but -0.15 < [Fe/H] < 0.15, 40 stars
 \item (Col. 10) Liu, et al. (2020), Figs. 4, 5, \& 6, 68 stars
 \item (Col. 11) Brewer, et al. (2016), 160 stars, CQ fit; age < 9.0 Gyr, 5400K < \Te < 6400K, 
 4.24 < log($g$), and -0.15 < $[Fe/H]$ < 0.15.
 \item (Col. 12) Brewer, et al. (2016), 160 stars, UL fit, same restrictions as CQ fit.  
 \end{itemize} 
 \end{threeparttable}
 \end{table*}

 
 In Sec.~\ref{sec:twins} below, we consider the question of whether the abundance
 deviations from solar values are correlated with elemental condensation
 temperatures,  \Tc\ \citep{lod03}.  Many studies have looked for such correlations (see
 references in Sec.~\ref{sec:data}).
 
 The chemical elements with stellar abundances
 have an obvious correlation of $[El/Fe]$ with GCE \citep{adib14}. 
 Basically, the volatile
 elements carbon and oxygen are more abundant relative to iron, silicon and
 other refractories in older stars.  Thus one should correct a star's abundance
 for GCE prior to looking for a correlation with \Tc: 
 The first part of Eq.~\ref{eq:aabb} describes the UL fit, while the second
 gives tha CQ fit, where the correction vanishes for 4.6 Gyr.
 \begin{eqnarray}
 [El/Fe]_{\rm corr} & = & [El/Fe]_{\rm obs}- [m*{\rm age} + b ], {\rm or} \nonumber \\
                    & = & [El/Fe]_{\rm obs}-  [aa\cdot ({\rm age}-4.6) +  \nonumber \\
                    &   &  bb\cdot ({\rm age}-4.6)^2].
\label{eq:aabb}
\end{eqnarray}               
\noindent The constants $m$, $b$, $aa$, and $bb$ belong to each element.  

But these GCE coefficients have additional meaning.  They give the mean
abundance for the sample as a function of age.  
A motion picture display may be seen in the iPoster from              
the 235th meeting of the American Astronomical Society
\footnote
{https://aas235-aas.ipostersessions.com/?s=58-BE-9F-2E-6D-57-C6-85-88-1E-B6-1D-29-37-EC-8E}
that shows the average abundance patterns $[El/H]$ as a function of atomic number.
Ten epochs are shown in steps of 1 Gyr based on the BD18 UL parameters of their Tab. 3.
They clearly illustrate the transition from a positive overall slope (POS) in young
stars to a V-shaped configuration in the oldest stars. Those two basic shapes are
illustrated by the squares (and stars) in Fig.~\ref{fig:diffs}.  

A dominant feature of these plots is the migration of the triplet Sr, Y, Zr from
high values in young to low values in old stars.  While Sr, Y, Zr move downward with
age, the lighter refractories, Mg, Al, Si (black dots between Z = 11, Na, and 
Z = 16, S) move up slightly.  
This accounts for the ratio $[Y/Mg]$ used as a Galactic clock
\citep{dasil12,mag21}, but also shows that the behavior of these two elements
is closely related to their neighbors in Z. 

Most of the downward migration of the Sr, Y, Zr triplet takes place within 2 Gyr
of the Sun's age.  This could be related to the ``bi-modality''  of s-process 
production discussed by  \citet{kam21}, as well as the two populations
discussed by NIS. 

In the young stars, with POS-type slopes the elements Mg-Zn align nearly
perpendicular to the overall slope.  In the older stars, this line 
rotates slightly counter clockwise, nearly as a whole.  In the oldest stars, the
heaviest elements of this group, Cu and Zn, disconnect, and move upward. This is due
primarily to eleven stars that were purged from the original BD18
list of 79 stars but are included in the file BD18stars\_plots in the zenodo archive.

We now compare the differential abundances of individual stars
with the mean values computed from the CQ
GCE coefficients.  Fig.~\ref{fig:diffs} shows two examples--of relatively young
(HIP 101905) and old (HIP 15527) stars.  The star symbols represent the elemental 
differential abundances for individual stars.
Ordinates (y-values) for the squares are given by Eq.~\ref{eq:squares};
we show the $Z$-dependence of the GCE coefficients explicitly:

\begin{eqnarray}
y(Z, {\rm age})&  = &  <[El/Fe]>_{\rm ave}  \\ \nonumber 
               &  = & aa(Z)\cdot ({\rm age}-4.6)+bb(Z)\cdot({\rm age}-4.6)^2
\label{eq:squares}
\end{eqnarray}

The young-star abundances display the expected pattern--
POS, both observed (stars) and predicted (squares).
The older star shows the typical V-shape, due to the
strong drop for the intermediate s-process-dominated elements Sr, Y, and Zr. 
In Fig.~\ref{fig:diffs} (left), the iron group (3d) elements follow the means (squares).  
The alpha elements scatter from the means (squares):
C, O and S are about 0.04 dex low, while Mg, Si, and Ca are approximately fit.  
All of the neutron addition (nA) elements are significantly higher than the 
means. 

Abundances in the older star fit the means more closely.  Individual stars
have their own usually small departures from the mean abundances for their age.
Plots of this kind may be seen for all the BD18 stars in 
ObsandPrePlots in the zenodo archive.
It will be seen that deviations from the
average patterns are common, and occasionally as large as 0.1 dex.

\section{Solar twins, condensation, and engulfment}
\label{sec:twins}

\citet[][henceforth, MEL09]{mel09} showed that the Sun may not be the 
best chemical sample of its 
neighborhood.  Their Fig. 3 is a good illustration,   
where logarithmic differential abundances--Sun minus twins--are 
plotted against condensation temperature, \Tc.  This plot shows the
highly volatile element C (\Tc\ = 40K) with an
excess of about 0.05 dex.  The highly refractive Al (\Tc\ = 1641K)
has a depletion of about 0.03 dex.  The
extremes of about 0.08 dex, are comparable to the differences
of Tab.~\ref{tab:Abdifs}, excepting BD18 and NIS.  
The corresponding slope, $-5\cdot 10^{-5} {\rm dex\ deg}^{-1}$
is similar in magnitude to those of (suspected and) confirmed planet host stars
(see LIU Table 5, Column 2).
  
Numerous workers have sought an optimum sample of representative
stars for the solar neighborhood.
The topic was reviewed in detail by \citet{yana21},
who discuss new criteria based on Gaia data, geochronology,
and magnetic activity.

Our ``twins'' were selected with four parameters:
\Te, $\log{(g)}$, $[Fe/H]$, and age, 
taken from the original papers (BD18, NIS, LIU, BR, and DM).
A dozen stars were were chosen based on minimum Euclidian distances
from the Sun in these parameters.                                                             
Prior to calculating the distances,
the variables were normalized to fall between -1 and +1.  
We selected the 12 stars with the smallest distances 
in hyperspace from the Sun for each of the surveys.

Average values of $[El/Fe]$ for the nA elements from carbon to zinc were obtained. 
This average of
the logarithms is not the same as the logarithm of the (non-logarithmic)
average \citep[][see Eq. 1]{bd18}, but differs, typically, by one or two units in the 
third decimal, which is unimportant for our purposes.  Note that  
averaging logs rather than taking an average of antilogs has the advantage
of weakening the effect of outliers.

The BD18 twins are given in Tab.~\ref{tab:Dataset}; the remaining sets are
in Tab~\ref{tab:4twins}.


A number of very impressive abundance surveys are related to 
correlations of differential abundances with \Tc\ as an indication
of planetary engulfment (see references in Sec.~\ref{sec:data}).

The study by \citet{spin21} argues strongly that
engulfment is the most likely explanation for many such correlations.  However,
they present an  argument independent of \Tc, essentially based on the 
probability of abundance anomalies due to engulfment being more prominant in stars with
higher effective temperatures with thinner convection zones.

\begin{figure*}
	\includegraphics[width=6in]{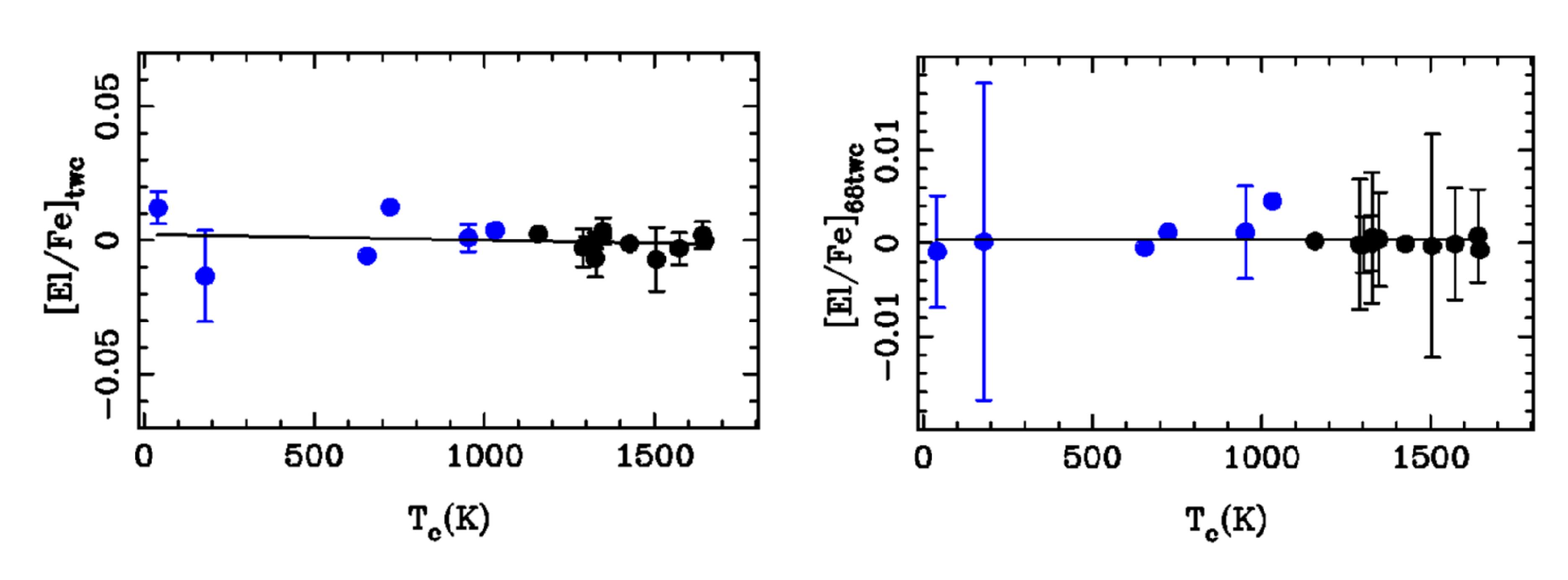}
   \caption{$[El/Fe]_{\rm twc\ or\ 68twc}$ = $[El/Fe]_{\rm Sun}
    -  [El/Fe]_{\rm Average\, twins}$. vs. condensation temperature, \Tc.
     Filled circles are BD18 points; grey (blue online) 
    points are volatiles C (40K), O (179K), Na (953K), S (655K), and Zn (723K)
   and Cu (1033K). 
   Black filled circles are refractories.  Error bars are the lengths of the
   standard deviations of BD18 minus NIS   
    Left: GCE CQ corrections are used. Right: UL GCE
    corrections used. 
    The straight lines are an unweighted least square fits.
    Note the different vertical scales, left and right. \label{fig:2Tcplots}}
\end{figure*}

Relative elemental abundance averages for each of the five surveys were
plotted as a function of condensation temperature, \Te\ \citep{lod03}.
Here, the relative abundances are for the Sun, with respect to the mean for the
twins of Tab.~\ref{tab:Dataset} and Tab.~\ref{tab:4twins}.
The average twin $[El/Fe]$ relative to the Sun and corrected with CQ for GCE
we designati $[El/Fe]_{\rm twc}$, for "twin, corrected":
\begin{eqnarray}
[El/Fe]_{\rm twc} & = & \frac{1}{12}\sum_{i=1}^{12}[El/H]_{\rm BD18}-[Fe/H]_{\rm BD18} \nonumber \\
                 & - &\left(aa\cdot({\rm age}-4.6)+ bb\cdot({\rm age}-4.6)^2\right)
\label{eq:twc}
\end{eqnarray}
\noindent 
The plots of Fig.~\ref{fig:2Tcplots} follow the orientation of MEL09, 
so that deficiencies of refractory elements
in the Sun result in a downward slope with increasing \Tc.

Only elements 
from carbon to zinc were used. 
We excluded the neutron-addition (nA) elements from consideration in the search
for \Tc\ correlations for reasons noted by \citet{mel14}, and adopted by \citet{cby20}
in their survey of the BD18 sample.
Many examples may be
seen in the plots of the BD18 abundances vs. \Tc, where the nA abundances 
scatter sometimes above (HIP 8507) and sometimes below (HIP 9349) the trends
for the lighter elements (see TccorGIF1 in the zenodo archive).
This is evidence that the neutron-addition elements have a 
distinct GCE from C-Zn, as argued by \citet{mel14}.

The BD18 survey has abundances for 18 elements, from C to Zn, plus 12 nA elements not
used here.  BR has one such, element, Y, and 14 from C to Ni, including the
ultravolatile, N, which is unavailable in the other two surveys.  DM has only the
intermediate volatiles Cu and Zn, and the refractories Mg - Fe.   

Twins for BR, DM, NIS, and LIU in Tab.~\ref{tab:4twins}
are listed in order of their distance in hyperspace
from the Sun.  Thus, of the BR twins, HD 9407 is closer to the
Sun in \Te, $log(g)$, $[Fe/H]$, and age than HD 157347.

\begin{table}  
	\centering
	\caption{Twins from four PDA surveys; 
    last row gives minimum and maximum ages in Gyr}
	\label{tab:4twins}
	\begin{tabular}{rrrr} 
    \hline
  BR(160*)&     DM(73*)  &      NIS(71*)&   LIU(68*)  \\
   HD     &      HD      &       HD     &      HD     \\
\hline
  9407    &      150437  &        2071  &     146233  \\  
  157347  &      29137   &        1461  &     140538  \\  
  152391  &      134987  &        4915  &     138004  \\  
  203030  &      115585  &         361  &     68168   \\  
  50692   &      90722   &        7134  &     10307   \\  
  147750  &      33822   &        6204  &     190406  \\  
  103828  &      109271  &       13724  &     106116  \\  
  85689   &      134664  &      196390  &     219542  \\
  157338  &      147513  &      189625  &     141937  \\  
  34411   &      76151   &      202628  &     159222  \\  
  24496   &      155968  &       12264  &     4915    \\  
  10307   &      189625  &       59967  &     217014  \\ \hline
  4.3--4.9 &     1.6--7.2  &      1.5--7   &  2.5--6.8  \\
\hline			  
\end{tabular}    
\end{table}      

Fig.~\ref{fig:2Tcplots} uses GCE-corrected abundances for 12 twins, by CQ (left) 
and 68 twins UL (right),
numerical values for the latter are from BD18's Tab.~\ 3.  Error bars are given
for elements in common for BD18 and NIS surveys.  The bar lengths are equal
to the standard deviations of $[El/H]_{\rm BD18} - [El/H]_{\rm NIS}$ for the 11 non neutron-addition elements
in common. S, Zn, Cu, and Mn (1158 K) were not determined in NIS.

Neither fit in Fig.~\ref{fig:2Tcplots} is significant\footnote{``Significance'' or ``sig''
is the probability that the correlation is due to chance.  If sig > 0.05,
the correlation is not considered significant.} The larger of the absolute value
of the two tiny slopes is $-0.21\cdot 10^{-5}$ for the 12 twins (left). 
This ``lack'' of an indication of depletion of the solar refractories was 
sufficiently surprising that a calculation was made using all 68 stars from BD18 as twins.
We again excluded 11 of their 79 stars.  The resulting plot, Fig.~\ref{fig:2Tcplots} (right)
supported the normality of the solar refractory abundances relative to these
two sets of twins. 

A further confirmation of the normality of solar refractories in BD18 
are the abundances of a subset
of 25 stars with ages within $\pm$ 1.5 Gyr of the Sun's age. Because of their
proximity to the Sun's age, GCE corrections are small.  Averages of $[El/Fe]$ are
very small from a minimum of $-$0.026 for carbon and a maximum of +0.017 for
calcium.  Values for the refractory Ti (1573 K) and Al (1641 K) were respectively
0.013 and 0.006.

 Scatter for the other surveys is larger than for BD18.  If we combine
the recent NIS and LIU into 26 points, unweighted linear fits (not shown) have slopes of
$-1.39 \times 10^{-5} {\rm dex\ deg}^{-1}$ (uncorrected for GCE), sig = 0.054, and
$-0.95 \times 10^{-5} {\rm dex\ deg}^{-1}$ (corrected), sig = 0.20, giving weak 
support to the depletion of solar refractories.  An examination of 60 NIS stars with
ages < 9 Gyr shows small depletion of all 10 elements from carbon to nickel.  
Depletions of refractories magnesium -- nickel are all < 0.02 dex if GCE
corrections are applied.  The greatest depletion is for oxygen, -0.04 dex.

It is possible to relate the slopes of abundance vs \Tc\ plots to an engulfed mass
using the toy model of \citet{cby21}.  The model gives the slope, $q$, of the
$[El/H]$ vs. \Tc plot.  Because $[El/H] = [El/Fe]-[Fe/H]$ there will be no difference
in the slope of plots of $[El/H]$ and $[El/Fe]$ for constant $[Fe/H]$.


We assume solar abundances \citep{asp21} for
the mass and composition of the convection zone (SCZ) as well as the composition
of the mass ($q\times M_{\rm BE}$) added.  Here, we take $M_{\rm SCZ} = 0.025 M_\odot$,
having the Sun's composition \citep{asp21}.  For negative slopes, the mass is added
to the average solar twin and missing from the Sun.

The added mass is $q$ times the earth's mass and has the earth's composition 
\citep{wn18}.  For other convection zone masses, the added mass must be adjusted
accordingly--e.g. in a linear regime, doubling the $M_{\rm SCZ}$ would require 
doubling the added mass.  

Quantitative values for the $q$ parameters and the corresponding slopes  are
given in Tab.~\ref{tab:qslope}.  The coefficient of $10^{-5}$ is fortuitously
nearly equal to $q$ because of the correlation of $M_{\rm BE}/M_{\rm SCZ}$ with
\Tc.

\begin{table}  
	\centering
	\caption{Slopes of $[El/H]$ for different values of $q$; 
    last row gives minimum and maximum ages in Gyr}
	\label{tab:qslope}
	\begin{tabular}{rrrr} 
    \hline
  $q$     &slope*10$^{5}$&  $q$         &slope*10$^{5}$  \\
\hline
  0.02    &   0.02       &    1.0       &    1.08     \\
  0.10    &   0.11       &    3.0       &    3.13     \\  
  0.33    &   0.36       &   10.0       &    9.35     \\  
  0.50    &   0.54       &   30.0       &   22.1      \\  \hline
\hline			  
\end{tabular}    
\end{table}      

The negative slopes of the abundances vs. \Tc\ indicate 0.21 to 1.4 $M_{\rm BE}$ was
added to the average solar twin but is missing from the Sun.  The tiny slopes of
Fig.~\ref{fig:2Tcplots} corresponds to no refractive mass difference in the
Sun and average twin.

\section{Summary}

We have examined precision differential abundances from five surveys.  We obtained
new GCE parameters from the BD18 set, using an algorithm that gives zero correction
for stars with solar age.  The new constrained quadratic (CQ) algorithm allows for 
curvature that is in agreement
for a number of elements with previous studies and in particular supports the conclusion
of NIS of two populations with different abundance patterns for stars younger and older than the Sun.  
For most elements, curvature described by our CQ fits was small; the overall agreement
BD18 parameters is good.
 
Sets of twelve stars were chosen from each of five surveys of F, G, and K stars as
solar analogues or ``twins.''  
The choice was based on proximity in a hyperspace consisting of
\Te, $\log{(g)}$, [Fe/H], and age.  Additionally, we considered a subgroup
of 68 BD18 stars as twins.

Similar parameters for the NIS, LIU, and the
larger DM and BR were obtained.   Results depend sensitively
on the selection of subsets of stars from the samples. 
Abundances from each set were averaged, and plotted
against condensation temperature.  The peak-to-peak scatter is nearly 0.1 dex,
In the case of the BD18 data, that scatter is about 0.025 dex.
The BD18 data does not support the thesis that the Sun is depleted in
refractory elements.  Examination of several subsets of the data shows generally
very small depletions of solar refractories but also depletions of oxygen.

Depletions of the refractories in the Sun are probably < 0.02 dex,
but depend on the choise of comparison stars (twins).

A simple model of addition of bulk earth material to a solar convection zone connects
the slopes of $[El/Fe]$ vs. \Tc\ plots to the amount of ingested (or missing)
refractory-rich material.  We estimate the amount of ingested (or missing) 
material is between zero and 1.4 earth masses.   


\section*{Acknowledgements}
This research has made use of the NASA Exoplanet Archive, which is operated by the 
California Institute of Technology, under contract with the National Aeronautics and 
Space Administration under the Exoplanet Exploration Program.  We also used the SIMBAD database
operated at CDS, Strasbourg, France \citep{wen00}.  Thanks are due to D. J. Bord
and to the referee for many useful comments and suggestions.
\vspace{0.2in}

\noindent C. Cowley @ https://orcid.org/0000-0001-9837-3662
\newline K. Y\"{u}ce @ https://orcid.org/0000-0003-1910-3344


\section{Data Availability}

This work is based entirely on 
data that may be obtained from the
CDS in Strasbourg France.



\appendix



\bsp	
\label{lastpage}
\end{document}